\documentclass{ws-procs975x65}
\begin{document}
\title{Active Star-forming galaxies in pairs in the 2dF}
\author{G. Sorrentino, B. Kelm, P. Focardi}
\address{Dipartimento di Astronomia, Via Ranzani 1, 40127 Bologna, Italy}
\maketitle
\abstracts{We explore the close environment of active star-forming (ASF) 
galaxies in the 2dF. We find ASF galaxies to be the most likely 
population to inhabit extremely close pairs located in 
low density regions. In addition, we find that  
ASF galaxies in these pairs are almost entirely concentrated in 
the [$-20$$\leq$M$_B$$\leq$$-19$] magnitude range. }
\section{Introduction}
The environment of a galaxy may have a significant effect on its 
spectral characteristics (Balogh et al.\,2003\,astro-ph/0311379). 
In a cluster the role of singular encounters has a negligible 
impact, but it is important to understand whether close approaches in 
low density environments are important, as the majority of galaxies are found 
in a group-like environment. 
Close encounters are predicted to trigger star-formation 
and eventually AGN activity.  
We investigate whether any excess of active star-forming galaxies 
is found in small systems identified in the 2dF (Colless et al. 2003). 
The 2dF sub-sample we examine includes 10695 galaxies 
with m$_B$ between 17 and 17.5. 
For each galaxy close neighbours (with 17$\leq$m$_B$$\leq$19.5) 
have been identified lying within a circular area of 0.25\,$h^{-1}$Mpc 
projected radius and $\pm$1000\,$km/s$ from the galaxy. 
To each galaxy a surface density parameter $\sigma$$_{0.25}$ has been 
assigned, which is the ratio between the number of neighbours 
and the area of the region enclosing its most distant neighbour. 
\begin{figure}[ht]
\centerline{\epsfxsize=5cm \epsfbox{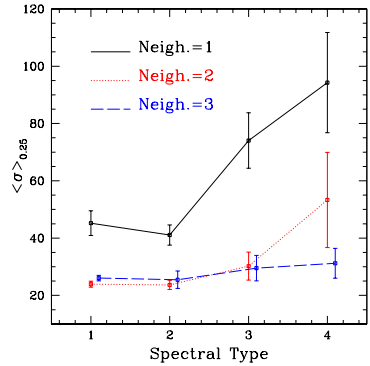}}   
\caption{Average value of the surface density parameter $\sigma$$_{0.25}$ for different spectral type galaxies. Sub-samples of galaxies with 1, 2 and 3 
neighbours are plotted separately. The standard error on the mean is shown.\label{inter}}
\end{figure}
\begin{figure}[ht]
\centerline{\epsfxsize=6cm  \epsfbox{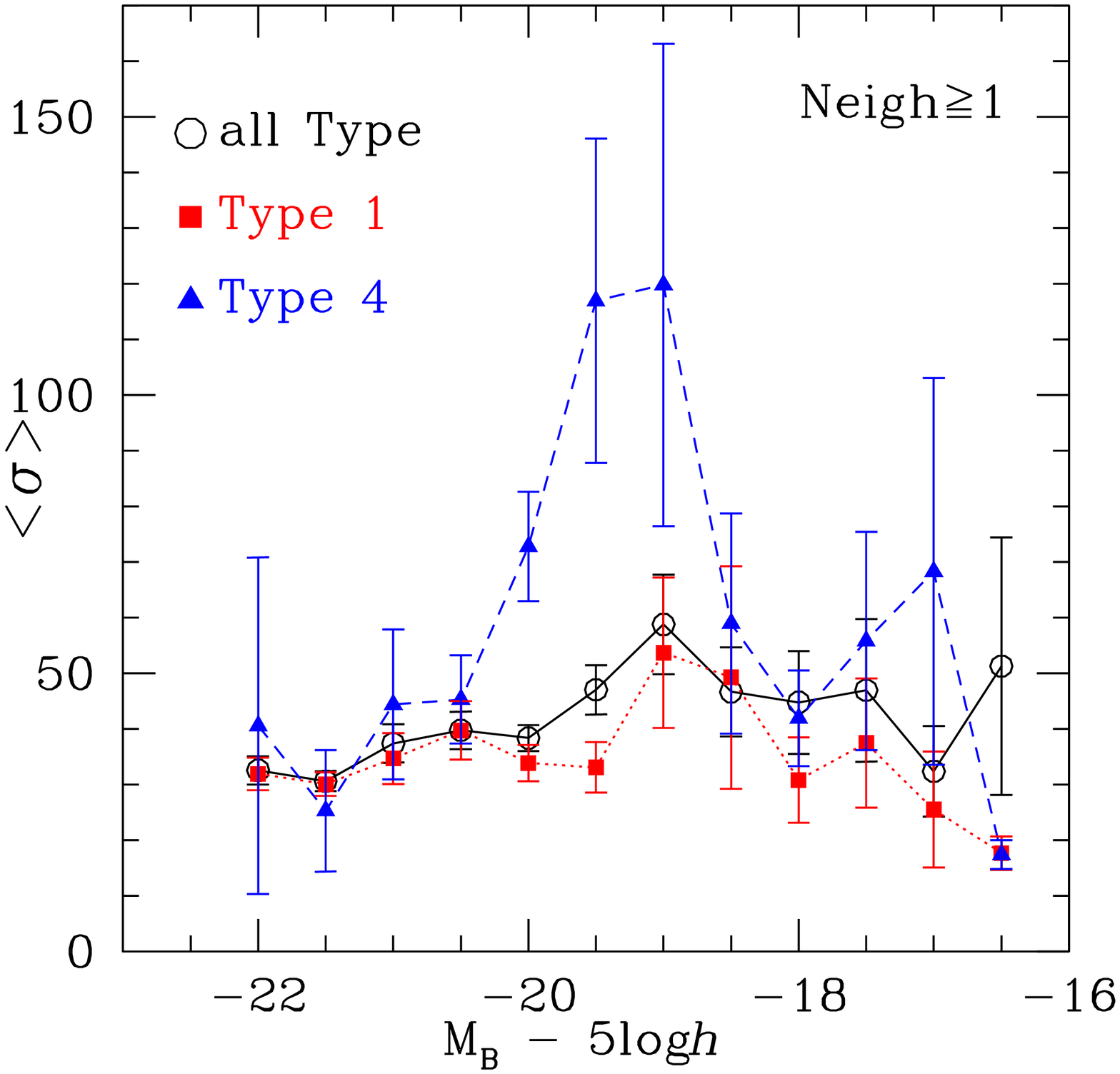}\epsfxsize=6cm \epsfbox{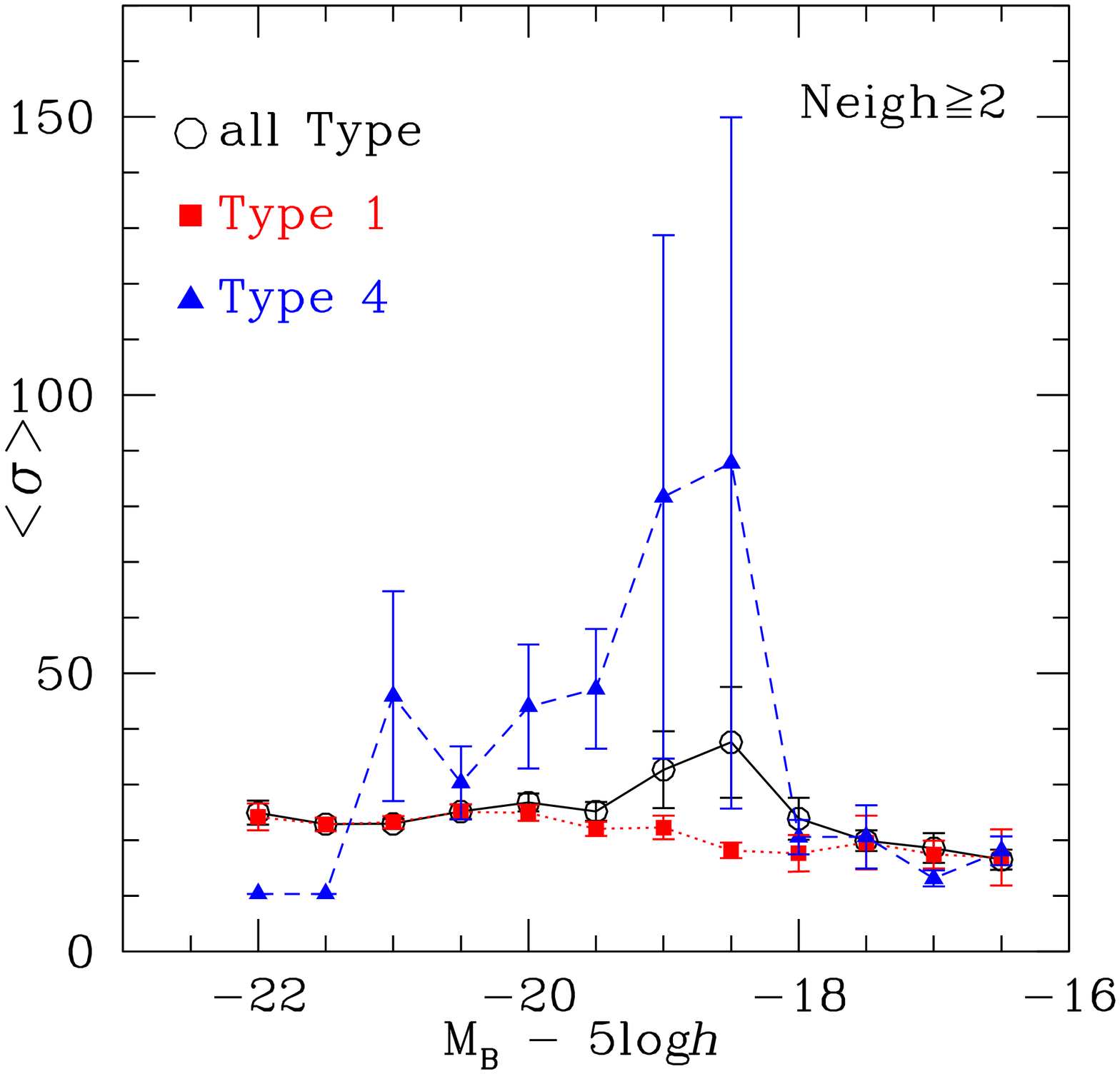}}      
\caption{Surface density parameter $\sigma$$_{0.25}$ (average value per 1 magnitude bin) as a function of M$_B$ for all galaxies and for T\,1 and T\,4 (ASF) galaxies. \label{inter}}
\end{figure}
\section{Active Star Forming galaxies in pairs}
In Fig.\,1 we show the average value of $\sigma$$_{0.25}$ 
vs. spectral type for galaxies in our sample. 
Spectral type is assigned as in Madgwick et al. (2002, MNRAS\,333,\,133). 
Galaxies in pairs, triplets and quartets are shown. 
It can be seen that the average value of $\sigma$$_{0.25}$ of ASF galaxies 
(T\,4) is significantly higher than those of non active galaxies 
(T\,1 and T\,2) in the pair sample, while no significant 
differences are seen in triplets and quartets. 
ASF galaxies appear therefore much more likely than 
non-active galaxies to inhabit extremely close pairs located in 
low density regions. High $\sigma$$_{0.25}$ values   
derived for ASF galaxies in pairs confirm that it is the distance to the 
nearest neighbour that is relevant (Lambas et al.,\,2002,\,astro-ph/0212222), 
and additionally, that the triggering efficiency rises  
in more isolated pairs.    
In Fig.\,2 we show $\sigma$$_{0.25}$ as a function of absolute magnitude 
for all galaxies in our sample and for ASF and T\,1 galaxies separately. 
The left panel shows $\sigma$$_{0.25}$ mean values derived for all galaxies 
displaying 1 or more neighbours (Neigh$\geq$1), the right panel shows values 
obtained when excluding pairs (Neigh$\geq$2). 
High $\sigma$$_{0.25}$ values associated to ASF galaxies 
are seen in the sample including pairs, while ASF galaxies in denser  
systems do show a less significant excess.      
Figure 2 also reveals that ASF galaxies with high values of $\sigma$$_{0.25}$ 
concentrate in a narrow magnitude range [$-20$$\leq$M$_B$$\leq$$-19$]. 
In this range ASF galaxies constitute between 7\% and 15\% of the 
whole galaxy population.   
Below M$_B$$\simeq$$-20$ the values of $\sigma$$_{0.25}$ are broadly similar between T\,1 and T\,4 galaxies  
indicating that, unlike the number of large scale neighbours (Kelm et al.\,2003\,astro-ph/0309268), the number of close neighbours does not discriminate between bright T\,1 and bright T\,4 galaxies.   
%
%
%
%
%
\end{document}